\documentclass[aps,pre,showpacs,superscriptaddress,twocolumn,floatfix]{revtex4}
\usepackage{graphicx}
\def\beq{\begin{equation}}
\def\eeq{\end{equation}}
\def\bea{\begin{eqnarray}}
\def\eea{\end{eqnarray}}
\begin{document}

\title{Persistence of a Rouse polymer chain under transverse shear flow} 

\author{Somnath Bhattacharya}

%\email{somnath@phy.iitb.ac.in}

\affiliation{Department of Physics, Indian Institute of Technology, Bombay, Powai, Mumbai-400 076, India.}

\author{Dibyendu Das}

%\email{dibyendu@phy.iitb.ac.in}

\affiliation{Department of Physics, Indian Institute of Technology, Bombay, Powai, Mumbai-400 076, India.}

\author{Satya N. Majumdar}

%\email{satya.majumdar@lptms.u-psud.fr}

\affiliation{Laboratoire de Physique Th\'eorique et Mod\`eles Statistiques, Universit\'e Paris-Sud, B\^atiment 100, 91405 Orsay Cedex, France.}

\date{\today}

\begin{abstract}
We consider a single Rouse polymer chain in two dimensions in presence of a transverse
shear flow along the $x$ direction and calculate the persistence probability 
$P_0(t)$ that the $x$ coordinate of a bead in the bulk of the chain does
not return to its initial position up to time $t$. We show that the persistence
decays at late times as a power law, $P_0(t)\sim t^{-\theta}$ with a
nontrivial exponent $\theta$. The analytical estimate of $\theta=0.359...$
obtained using an independent interval approximation is in excellent agreement
with the numerical value $\theta\approx 0.360\pm 0.001$.
\end{abstract}

%\keywords{polymer, shear,....}

%{polymer solutions,probability - stochastic process}

\maketitle
%\begin{multicols}{2}

\section{Introduction}

Polymer dynamics plays a central role in material science
and biology.  In particular, dynamics of an individual flexible or
semi-flexible polymer under a suitable shear force has been of great
interest~\cite{gennes,LeDuc,smith,doyle,chu,chertkov,gera}. Shear
force comes into play when a fluid flows past a surface.  Substantial
effort has been undertaken to investigate the motion of polymers in
shear field.  Previously, studies were done on bulk samples using
light-scattering and birefringence experiments. Recently, the dynamics
of a single polymer has also been investigated using video-microscopy.
Under a shear stress, such a polymer shows tumbling in addition to a
longitudinal stretching~\cite{LeDuc,smith}.  If one takes a tethered
polymer, whose one point is made immobile, then tumbling leads to a
cyclic motion of the spatially constrained polymer about a mean
position~\cite{doyle}. Also statistics of polymer orientation angles
has been of interest~\cite{chertkov,gera}. These properties studied
both experimentally and theoretically can be classified as long-time
transport phenomena. In contrast, in this paper we explore the persistence
or the survival
probability behaviour of a flexible polymer
chain under transverse shear flow within the paradigm of the
simple Rouse model where the polymer chain consists of beads
or monomers connected by harmonic springs~\cite{Rouse}.
We will show that even in this simple model, the 
persistence at late times decays as a power law characterized
by a nontrivial exponent.

The survival/persistence probability $P_0(t)$ that a stochastic process
$X(t)$ does not cross zero upto time $t$ is a quantity of long
standing interest in probability theory and with many practical
applications~\cite{BL}. The derivative $F(t) = -dP_0(t)/dt$ is the
first-passage probability~\cite{Redner}. In many nonequilibrium
many body systems, the persistence has been found to decay as a power law at
late times, $P_0(t)\sim t^{-\theta}$. The exponent $\theta$ is called
the persistence exponent and has been a subject of much theoretical,
numerical and experimental studies in recent times~\cite{Review}. The
exponent $\theta$ is often nontrivial and is generally hard to calculate 
analytically even in
simple systems such as the linear diffusion equation starting from
random initial conditions~\cite{satya}. The reason for this
difficulty can be traced back to the fact that the spatial
interactions in these extended systems makes the local stochastic
field $X(t)$ a `non-Markovian' process in time~\cite{Review}.

In this paper we study the persistence properties of a Rouse chain in
$2$-dimensions in presence of a transverse shear velocity field which
is non-random.  We show that the persistence probability in this
system decays at late times as a power law with a nontrivial
persistence exponent $\theta\approx 0.36$ that we compute numerically
as well as analytically within an independent interval approximation
(IIA). We note, that the current problem is in contrast to similar
problems considered in ``random'' flow fields earlier.  For example,
for a Rouse chain~\cite{Rouse} of infinite length, the transport
properties~\cite{OB,WL,JOB} and the persistence properties~\cite{SM}
in a quenched random velocity flow field have been studied.

The paper is organized as follows. In Section II, we define the model 
precisely and summarize our main results.
In Section-III we present exact calculations of the two-time correlation
functions in our model. These results are used next in Section IV to
calculate the persistence exponent analytically within the IIA.
Sections V and VI describe details
of the numerical methods and finally we conclude in Section VII. 
 
\section{The model and main results}

We consider a Rouse polymer chain embedded in a $2$-dimensional plane. The chain consists
of beads connected by harmonic springs~\cite{Rouse}. In addition, the chain 
is advected by shear velocity flow field. Let $[x_n(t), y_n(t)]$ denote
the coordinates of the $n$-th bead at time $t$ which evolve
with time according to the following equations of motion
\begin{eqnarray}
\frac{dy_n}{dt} &=& \Gamma\left(y_{n+1}+y_{n-1}-2\,y_n\right) + \eta_1(n,t) \label{evoly1} \\
\frac{dx_n}{dt} &=& \Gamma\left(x_{n+1}+x_{n-1}-2\,x_n\right)+ v \left(y_n(t)\right) + \eta_2(n,t),
\label{evolx1}
\end{eqnarray}
where $\Gamma$ denotes the strength of the harmonic interaction between nearest neighbour beads,
$\eta_1(n,t)$ and $\eta_2(n,t)$ represent the thermal white noises along the $y$ and $x$
directions respectively that are uncorrelated.
The transverse shear velocity field $v(y)$ is linear  
\begin{eqnarray}
v(y) = y. 
\label{qn1}
\end{eqnarray}
For a finite chain with $N$ beads, Eqs. (\ref{evoly1}) and (\ref{evolx1}) are 
valid only for the $(N-2)$ interior beads. The two boundary beads 
will have slightly different equations of motion.
However, for an infinitely large chain ($N\to \infty$), the translational
invariance along the length of the chain is restored since the  
boundary conditions become irrelevant for late time dynamics. Since we are 
mostly interested in the late time properties, one can
make further simplifications by replacing the discrete index $n$ of the beads
by a continuous variable $s$ and subsequently replace the discrete Laplacian by a
continuous second derivative along the $s$ direction. The coarse
grained versions of the evolution equations (\ref{evolx1}) then become 
\begin{eqnarray} 
\label{yeq}
\frac{\partial y(s,t)}{\partial t} &=&\Gamma \frac{\partial^2 y(s,t)}{\partial s^2} + 
\eta_1(s,t),\\
\label{xeq}
\frac{\partial x(s,t)}{\partial t} &=& \Gamma \frac{\partial^2 x(s,t)}{\partial s^2} + y(s,t). 
%+ \eta_2(s,t).
\end{eqnarray} 
Note that we have also dropped 
the $\eta_2(s,t)$ term in the second equation. This is simply
because one can easily show that the noise term $\eta_2(s,t)$ becomes
insignificant compared to the shear force term $y(s,t)$ at late times.
Hence for late time asymptotic properties we can ignore the noise $\eta_2(s,t)$.

In the absence of harmonic interactions ($\Gamma=0$), the beads become
independent and the coordinates of any (say the $n$-th) bead 
represents a two-dimensional Brownian walker in a shear flow\cite{KR}. 
Equivalently , in this limit, the $x$ coordinate of the walker
evolves as $d^2x_n/dt^2=\eta_1(n,t)$, i.e., it represents
a randomly accelerated particle. The persistence probability
of the $x$-coordinate, i.e., the probability that the
$x$ coordinate does not cross zero up to time $t$ 
is known to decay as $\sim t^{-{1/4}}$~\cite{accl}.  
Recently, the persistence of a single random walker for
various other deterministic velocity functions $v(y)$ has also been 
studied~\cite{gonos,bm}. 
Interestingly it has been shown that for all
odd functions $v(y)$ survival probability decays as $t^{-{1/4}}$~\cite{gonos}. It
turns out that the same $t^{-1/4}$ decay also holds in the case when
$v(y)$ is not a deterministic function, but represents a 
quenched random transverse velocity field with short-range correlations~\cite{redner,satya1}.
This model of a single random walker in presence of a random transverse velocity field
is known as the Matheron-de-Marsily model\cite{mdm} whose transport properties had 
been studied earlier extensively\cite{bg}, but the studies of persistence properties
are relatively new\cite{redner,satya1,soumen}.  

In this paper we study the persistence probability of the
$x$ coordinate of the $n$-th bead in the presence of harmonic interaction
$\Gamma\ne 0$. Due to the translational invariance along the
length of the chain in the bulk, the persistence probability is 
independent of the label $n$ of the bead. 
We also absorb the factor $\Gamma$ by properly rescaling the time.
Note that the continuum equation (\ref{yeq}) for the $y$ coordinate
is precisely the Edwards-Wilkinson equation of one dimensional
interface~\cite{EW} and its persistence properties are known, both 
theoretically~\cite{krug,constantin} and also experimentally~\cite{expint}.
Here we focus on the $x$ coordinate and define the persistence 
as follows
\begin{eqnarray}
P_0(t) &=& {\rm {Prob}} [x(s,t')\ne x(s,0)\,\, \nonumber \\ 
&& {\rm for\,\, all\,\,} t': \,\, 0\le t'\le t ], 
\label{perx1}
\end{eqnarray}
i.e., $P_0(t)$ is the probability 
that the $x$ coordinate of any bead 
does not return to its initial position within the 
time interval $[0,t]$. 

The initial conditions for the chain coordinates do not play
any role in the persistence probability. This is due to the
fact that the evolution equations are linear, so we can
redefine the change in positions $x(s,t)-x(s,0)$ and $y(s,t)-y(s,0)$
as the relevant coordinates which satisfy the same evolutions equations.
Hence, for the evolution equations (\ref{yeq}) and (\ref{xeq}) we can 
set the initial conditions $x(s,0)=0$ and $y(s,0)=0$ without 
any loss of generalities.  

Our main results can be summarized as follows.  
We show that $P_0(t)\sim t^{-\theta}$ at late times $t$
where the persistence exponent $\theta$ 
has a nontrivial value. The numerical value
$\theta\simeq 0.360\pm 0.001$,
%, obtained by two different methods,
is in excellent agreement with the analytical value $\theta=0.359...$
obtained within the IIA method. Thus as one switches on the harmonic interaction 
$\Gamma\ne 0$ between the beads, the exponent $\theta\simeq 0.36$ increases
from its value $\theta=1/4$ for $\Gamma=0$. Thus the $x$ coordinate
of a bead survives less in presence of harmonic interactions, i.e.,
the interaction enhances the return probability.

\section{Calculation of exact two-time correlation functions}

The stochastic processes $x(s,t)$ and $y(s,t)$ evolving via Eqs. (\ref{yeq}) and (\ref{xeq})
are both Gaussian at late times since the evolution equations are linear.
A Gaussian process is completely specified by its two-time correlation function.
More detailed quantities such as the persistence probability, in principle,
is a complicated functional of the two-time correlation function.   
In this section we compute the two-time correlation functions
exactly and use these functions later for computing the
persistence probability in Section IV.

To begin with, we Fourier transform 
Eq. \ref{yeq}. We define $\tilde{y}(k,t) = \int_{-\infty}^{+\infty} y(s,t) 
\exp(-isk) ds$, and $\tilde{\eta}_1(k,t) = \int_{-\infty}^{+\infty} 
\eta_1(s,t) \exp(-isk) ds$. This implies, 
\begin{equation}
\label{ytildeq}
\frac{\partial{\tilde{y}(k,t)}}{\partial{t}}=-k^2\tilde{y}(k,t) + \tilde{\eta_1}(k,t)
\end{equation}
Assuming flat initial condition (i.e., $y(s,t)=0$), Eq. (\ref{ytildeq}) gives 
\begin{equation}
\label{ht1s}
\tilde{y}(k,t) = \exp(-k^2 t) \int_{0}^{t} \tilde{\eta_1}(k,t') \exp(k^2 t') dt',\\ 
\end{equation}
which in turn implies the correlation function 
\bea
&& \langle \tilde{y}(k_1,t')\tilde{y}(k_2,t'')\rangle  =  \nonumber \\ 
&& \frac{\delta(k_1 +k_2)}{2{{k_1}^2}} [\exp(-{k_1}^{2}|t'-t''|) - \exp(-{k_1}^{2}(t'+t''))]. \nonumber \\
\label{ycorr}
\eea
 
For Eq. (\ref{xeq}), defining $\tilde{x}(k,t) = \int_{-\infty}^{+\infty} x(s,t) 
\exp(-isk) ds$ and again considering flat initial condition (i.e., $x(s,0)=0$),
we get 
\begin{equation} 
\label{ht2s}
\tilde{x}(k,t) = \exp(-k^2 t) \int_{0}^{t} \tilde{y}(k,t') \exp(k^2 t') dt',
\end{equation}
which further implies, 
\begin{eqnarray}
\label{xcorr}
&&\langle \tilde{x}(k_1,t_1)\tilde{x}(k_2,t_2)\rangle = 
\exp(-{k_1}^{2}t_{1}-{k_2}^{2}{t_2}) \times \nonumber \\ 
&& \int_{0}^{t_1} dt' \int_{0}^{t_2} dt'' 
\langle\tilde{y}(k_1,t')\tilde{y}(k_2,t'')\rangle\, \exp(k_1^2 t' + k_2^2 t''). \nonumber \\ 
\end{eqnarray}
Substituting Eq. (\ref{ycorr}) in Eq. (\ref{xcorr}), we get
\begin{eqnarray}
&&\langle \tilde{x}(k_1,t_1)\tilde{x}(k_2,t_2)\rangle \nonumber \\
%&=&\exp(-{k_1}^{2}(t_{1}+{t_2})) \delta(k_1 +k_2) \times \nonumber \\ 
%&& \int_{0}^{t_1} dt' \int_{0}^{t_2} dt''\, \frac{[\exp({k_1}^{2}(t'+t'' - |t' - 
%t''|))-1]}{2{k_1}^{2}} \nonumber \\
&=&   \delta(k_1 +k_2) \int_{0}^{t_1} dt' \int_{0}^{t_2} dt''\times \nonumber \\ 
&& \frac{\exp(-{k_1}^{2}(t_1+t_2-2~ {\rm min}(t',t'')))-\exp(-{k_1}^2(t_1+t_2))}{2{k_1}^{2}}. 
\nonumber \\
\end{eqnarray}
By inverting the Fourier transform above, we obtain the correlation
function 
\begin{eqnarray}
\label{at1t2}
&&C(t_1,t_2) = \langle x(s,t_1) x(s,t_2)\rangle \nonumber \\
&=& \int_{0}^{t_1} dt' \int_{0}^{t_2} dt'' \int_{-\infty}^{+\infty} dk_{1} \int_{-\infty}^{+\infty} dk_{2} \nonumber \\
&&\exp(i(k_1+k_2)s) ~ \langle \tilde{x}(k_1,t^{'})\tilde{x}(k_2,t^{''})\rangle \nonumber \\
&=& \int_{0}^{t_1} dt^{'} \int_{0}^{t_2} dt^{''} \int_{-\infty}^{+\infty} dk_1 
\nonumber \\
&&\left[\frac{1-\exp(-{k_1}^{2}(t_1+t_2))}{2{k_1}^2} - \frac{1-\exp(-{k_1}^{2}(t_1+t_2 - \tilde{t}))}{2{k_1}^2} \right], \nonumber \\
\end{eqnarray}
where $\tilde{t} = 2~ {\rm min}(t',t'')$. After some algebra, Eq. 
(\ref{at1t2}) leads to,
\begin{eqnarray}
&& C(t_1,t_2)  = \nonumber \\ 
&& B \left[ t_{1}t_{2}\sqrt{t_1+t_2} - \frac{1}{5} \{ (t_1+t_2)^{5/2} + |t_2 -t_1|^{5/2} \} \right], \nonumber \\
 \label{at1t2a}
\end{eqnarray}
where $B$ is an unimportant constant.

Note that due to the translational invariance in the bulk, the correlator
of the process $x(s,t)$ does not depend on the location $s$ of the bead along the chain.
Thus, for simplicity of notations, we can now drop the label $s$ and
consider $x(t)$ as the relevant Gaussian process with the correlator
$C(t_1,t_2)=\langle x(t_1) x(t_2)\rangle$ as given in Eq. (\ref{at1t2a}).
Clearly the process $x(t)$ is non-stationary since
its two-time correlator in Eq. (\ref{at1t2a}) depends
on both $t_1$ and $t_2$ and not just on
their difference. One can however define 
a logarithmic time $T=\ln t$ and consider the normalized process 
$X(T)= x(t)/{\sqrt {\langle x(t)^2\rangle}}$ in $T$~\cite{satya2}.
The survival or no zero crossing probability is clearly the same for
both the normalized process $X(T)$ and the original unnormalized processs $x(t)$.
It then follows from Eq. (\ref{at1t2a}) that the autocorrelation function
of this normalized Gaussian process $X(T)$ is stationary in the $T$
variable and is given by 
\begin{eqnarray}
\label{nc0}
A (T) & = & \frac{C(t_1,t_2)}{\sqrt{C(t_1,t_1)C(t_2,t_2)}}, \nonumber \\
%& = & \frac{5}{\sqrt{2}} \left(\frac{t_2}{t_1}\right)^{5/4}\left[\frac{t_1}{t_2}\sqrt{1+\frac{t_1}{t_2}} - \frac{1}{5} \left(1+\frac{t_1}{t_2}\right)^{5/2} + \frac{1}{5} \left(1-\frac{t_1}{t_2}\right)^{5/2}\right], \nonumber \\
& = & \frac{5}{\sqrt{2}} \exp(\frac{5}{4}T)~ [\exp(-T)\sqrt{1+\exp(-T)} \nonumber \\ 
&& - \frac{1}{5} \left(1+ \exp(-T) \right)^{5/2} + \frac{1}{5} \left(1-\exp(-T) \right)^{5/2}]. 
\nonumber \\
\end{eqnarray}
This form of the stationary autocorrelator will be used in the next section
to compute the persistence probability.

\section{Calculation of the persistence exponent $\theta$}

We have thus mapped our problem to a Gaussian stationary process in $T=\ln t$ variable with a
prescribed correlator $A (T)$ and we want
to calculate the probability $P_0(T)$ that the process does not cross
zero up to time $T$. For a general correlator $A(T)$, the computation of
$P_0(T)$ is very hard~\cite{slepian,BL,Review}. However, some  
general results are known for the late time behavior of $P_0(T)$. For example, it is 
known~\cite{slepian,BL,Review} that when $A(T)$ decays
faster than $1/T$ for large $T$, the persistence probability $P_0(T)$
decays exponentially, $P_0(T)\sim \exp(-\theta T)$. 
Since, in our case, $A(T)$ is Eq. (\ref{nc0}) decays faster than $1/T$ for large $T$,
we expect $P_0(T)\sim  \exp(-\theta T)$.
In terms of the original time
variable, $t=e^T$, this would signify a power law decay of the persistence
$P_0(t)\sim t^{-\theta}$ for large $t$. Thus the inverse decay rate $\theta$ in
the $T$ variable 
is precisely the exponent of the algebraic decay in the real time $t$.

While we were not able to compute the exponent $\theta$ exactly, one
can obtain a very accurate analytical estimate of $\theta$ using
the IIA method that was first 
used in the context of persistence in diffusion equation~\cite{satya}.
This method works reasonably well only for {\it smooth} Gaussian stationary processes.
A process is smooth if $A(T)=1- a T^2+\ldots$ for small $T$. In that case, the
process has a finite mean density $\rho=\sqrt{-A''(0)}/\pi$~\cite{Rice} of zero crossings.
For our process, the correlator in Eq. (\ref{nc0}) can be expanded for small $T$
\bea
A(T \to 0) = 1 - \frac{15}{16}T^2 + \frac{1}{\sqrt{2}} T^{5/2},
\eea
indicating $a=15/16$ and thus proving that the process is smooth.

In the IIA, applicable only to smooth processes, one assumes that the 
intervals between successive zero crossings of a
Gaussian stationary process are statistically independent. 
Within this
approximation, one can then express 
the distribution $P(T)$ of the intervals between successive zero crossings 
in terms of the  
correlation
function $A(T)$ in the Laplace space~\cite{satya}
\bea
\tilde{P}(s) = \frac{1 - ({\langle T \rangle}/2)s[1 -
s \tilde{A}(s)]}{1 + ({\langle T \rangle}/2)s[1 - s \tilde{A}(s)]}.
\label{iia1}
\eea
Here $\tilde{P}(s)$ and $\tilde{A}(s)$ are the Laplace transforms
of $P(T)$ and $A(T)$ respectively and 
$\langle T \rangle= 1/\rho=\pi/\sqrt{-A''(0)}$ is the mean interval size. 

Using the exact expression of $A(T)$ from Eq. (\ref{nc0}), we can find
$P(T)$ from the above formula. The persistence probability
$P_0(T)$ is simply related to the interval distribution~\cite{satya},
$d^2P_0(T)/dT^2= P(T)/{\langle T\rangle}$. Since we expect
$P_0(T)$ to decay exponentially at late times $T$, i.e. 
$P_0(T)\sim \exp(-\theta T)$,
it follow that the interval distribution $P(T)$ will
also have the same late time decay, $P(T)\sim \exp(-\theta T)$
with identical exponent $\theta$.
This means that the Laplace transform
${\tilde P}(s)$ must have a simple pole at $s=-\theta$. In other words
the denominator in Eq. (\ref{iia1}) must have a root at
$s=-\theta$. Substituting $s=-\theta$, the denominator reads
\begin{equation}
\label{Gtheta}
G(\theta)= 1- (\theta/2)(\frac{\pi}{\sqrt{2 a}})\left[1+\theta ~\tilde{A}(-\theta)\right],
\end{equation}
where we have put $\langle T \rangle = {\pi}/{\sqrt{2 a}}$ with $a=15/16$
and $\theta$ is given by the smallest positive root of $G(\theta) = 0$. 

To determine the root of $G(\theta)=0$ accurately, it is convenient
to switch variables 
and define $x = exp(-T)$, such that 
\begin{equation}
\label{ax}
A(x)=\frac{x^{3/4}}{\sqrt{2(1+x)}}\left[2 + \frac{2x-1}{1+\sqrt{1-x^2}} -x + \sqrt{1-x^2} \right],
\end{equation}
and then Eq. (\ref{Gtheta}) becomes,
\bea
G(\theta) &=&  \frac{1}{\sqrt{2}} \int_0^1 dx \frac{x^{-(\theta+1/4)}}{1+x} \times \nonumber \\ 
&&\left[ 2-x +\sqrt{1-x^2} + \frac{2x-1}{1+\sqrt{1-x^2}} \right] = 0. \nonumber \\
\label{Gx=0}
\eea
Solving Eq. (\ref{Gx=0}) numerically gives 
\bea
\theta_{\rm IIA}=0.359....
\label{the0} 
\eea
Thus the persistence probability $P_0(t)\sim t^{-\theta}$ decays
algebraically for large time $t$ with a  
nontrivial exponent, whose analytical value within the IIA is $\theta_{\rm IIA}=0.359..$. 

\section{Simulation of discretised Langevin equations}

In this section we describe simulation of the Rouse chain evolving via 
Eqs. (\ref{evoly1}) and (\ref{evolx1}) and further discretised in time $t$ as 
\begin{eqnarray}
y_n(t_{m+1})&=& y_n(t_m) + {\Delta t}[ y_{n+1}(t_m)+y_{n-1}(t_m) \nonumber \\ 
&-& 2\,y_n(t_m)] + \sqrt{\Delta t}\,\zeta_1(n,t_m), 
\label{disy} \\ 
x_n(t_{m+1})&=&x_n(t_m)+ {\Delta t}[ x_{n+1}(t_m)+x_{n-1}(t_m) \nonumber \\
&-& 2\,x_n(t_m)] + {\Delta t} \, y_n(t_m), 
\label{disx} 
\end{eqnarray} 
where $t_m=m {\Delta t}$.  For the boundary points $n=1$ and $n=N$, we
use free boundary conditions, i.e., we hold $x_0=x_1$, $y_0=y_1$,
$x_N=x_{N+1}$ and $y_N=y_{N+1}$ for all times $t_m$.  We choose
$\Delta t = 0.1$ in our simulations \cite{krug}, and used chain
lengths of size $N = 1000-10000$. The variable $\zeta_1(n, t_m)$ is an
independent Gaussian variable for all $n$ and $t_m$ and 
distributed with zero mean and unit variance.

\begin{figure}
\includegraphics[width=6.0cm,angle=-90]{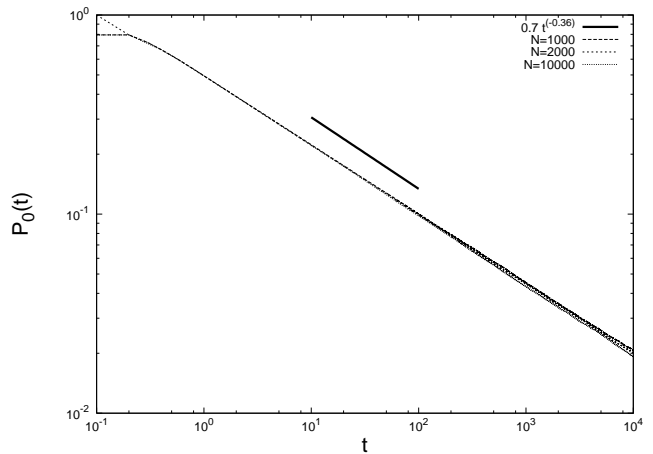}
\caption{Persistence probability $P_{0}(t)$ versus 
$t$, simulated for the polymer chain length $N=1000, 2000$ and $10000$.   
The fitted power law $t^{-0.360}$ is shown by plotting a thick line above 
the data.}
\label{numpersist0}
\end{figure}

The persistence probability for $x_n$ upto time $t_m$ was
obtained by keeping track of the fraction of $x_n$'s that have
$sgn[x_n(t_m)]$ same as $sgn[x_n(1)]$ for all times starting from $1$ to
$t_m$. The data are shown in fig.  \ref{numpersist0}.  Typically each
data curve in fig. \ref{numpersist0} was obtained by averaging over
$500$ thermal histories. We find that $P_0(t)$ decays as a
power law $\sim t^{-\theta}$ with $\theta \simeq 0.360 \pm
0.001$. The latter value is in good agreement with the IIA estimate 
in Eq. (\ref{the0}).

\section{Simulation of the Gaussian process}

Since a Gaussian stationary process $X(T)$ is completely specified by its 
stationary correlator $A(T)$, one can simulate the process by constructing a time-series
with the same correlator.
In the frequency domain (Fourier space) the
corresponding correlator is $\langle \tilde{X}(\omega_1)
\tilde{X}(\omega_2)\rangle = 2 \pi
\tilde{A}(\omega_1)\delta(\omega_1+\omega_2)$, where
${\tilde A}(\omega)=\int dt e^{i\omega T} A(T) $ is the 
Fourier transform of $A(T)$. The latter 
formula allows us to easily generate stochastic processes
\bea
\tilde{X}(\omega) = \tilde{\eta}(\omega) \sqrt{\tilde{A}_{S0}(\omega)},
\label{Xomega}
\eea 
where $\tilde{\eta}(\omega)$ is a Gaussian white noise with
$\langle \tilde{\eta}(\omega_1)\tilde{\eta}(\omega_2)\rangle = 2 \pi
\delta(\omega_1+\omega_2))$.

\begin{figure}
\includegraphics[width=6.0cm,angle=-90]{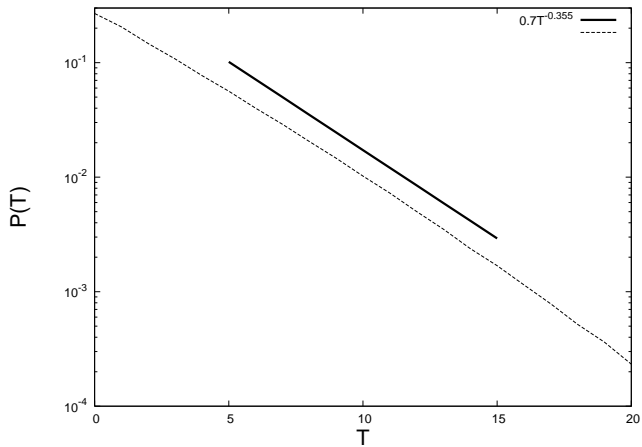}
\caption{The graph (dashed line) shows $P(T)$ vs. $T$ from the 
simulation of the Gaussian process $X(T)$. The upper short thick line 
goes as $0.7 \exp(-0.355 T) $.}
\label{gaus-zeroX}
\end{figure}

We performed simulations following the above route by first
constructing random functions $\tilde{X}(\omega)$ as per
Eq. (\ref{Xomega}) for discrete $\omega$'s. Then we did a discrete
inverse Fourier transform to obtain the times series $X(T)$
\cite{krug}. After generating $10^6$ such random time-series of
$X(T)$, we used them to calculate the probability density function
$P(T)$ of intervals between two consecutive zero-crossings. In the
calculation, the time-step size used was $\delta T=1$, and
$T_{max}=50$ as we found that $A(T)$ almost vanishes for $T > 50$. As
stated earlier, in terms of the variable $T$, both $P(T)$ and $P_0(T)$
decay as $\sim \exp(-\theta T)$. Hence, from the decay of $P(T)\sim
\exp(-\theta T)$, we estimated $\theta$. In Fig.  \ref{gaus-zeroX}, we
have shown $P(T)$ versus $T$, and we find the decay constant $\theta
\approx 0.355$. The latter value is slightly smaller than
the $\theta$ obtained from IIA and the Langevin simulation, because
the step-size $\delta T = 1$ was a bit large and we missed some intervals
smaller than that.

\section{Conclusion}

In summary, we have studied the persistence probability of the
$x$ coordinate of a bead in the bulk of a Rouse
polymer chain advected by a shear flow field. We have 
shown that the persistence probability decays 
as a power law in time at late times and the
associated persistence exponent $\theta\approx 0.36$ 
is nontrivial. We have computed this exponent analytically
within an independent interval approximation and also
determined it numerically by two different methods.
The analytical result is in excellent agreement with
the numerical simulations. 

There are several directions in which our work can be extended.
Here we have considered the Rouse chain embedded in two
spatial dimensions. It should be relatively straightforward to extend
our method to calculate the persistence properties of the
Rouse chain in higher dimensions in presence of
a transverse shear flow. It would also be of interest
to study the persistence properties of the polymer chain
in a more realistic setting going beyond the simple Rouse model, e.g,
in presence of excluded volume interactions. 

Acknowledgment: The authors acknowledge grant no. $3404-2$ of ``Indo-French
Center for the Promotion of advanced research (IFCPAR)/Centre
Franco-Indien pour la promotion de la recherche avancee (CEFIPRA)''
for financial support.

%\end{multicols}
 
\end{document}